\documentstyle[12pt]{article}
\begin{document}

\begin{titlepage}
\begin{center}
\today     \hfill    LBNL-40331 \\
~{} \hfill UCB-PTH-97/23 \\

\vskip .25in

{\large \bf New Mechanism of Flavor Symmetry Breaking\\
from Supersymmetric Strong Dynamics}\footnote{This work
was supported in part by the Director, Office of Energy Research, Office of
High Energy and Nuclear Physics, Division of High Energy Physics of the U.S.
Department of Energy under Contract DE-AC03-76SF00098.  LJH was also supported
in part by the National Science Foundation under grant PHY-95-14797.}

\vskip 0.3in

Christopher D. Carone$^1$, Lawrence J. Hall$^{1,2}$, and Takeo Moroi$^1$

\vskip 0.1in

{{}$^1$ \em Theoretical Physics Group\\
     Ernest Orlando Lawrence Berkeley National Laboratory\\
     University of California, Berkeley, California 94720}

\vskip 0.1in

{{}$^2$ \em Department of Physics\\
     University of California, Berkeley, California 94720}
        
\end{center}

\vskip .1in

\begin{abstract}
We present a class of supersymmetric models in which flavor symmetries 
are broken dynamically, by a set of composite flavon fields.  The strong 
dynamics that is responsible for confinement in the flavor sector also 
drives flavor symmetry breaking vacuum expectation values, as a consequence 
of a quantum-deformed moduli space.  Yukawa couplings result as a power 
series in the ratio of the confinement to Planck scale, and the 
fermion mass hierarchy depends on the differing number of preons in 
different flavor symmetry-breaking operators.  We present viable 
non-Abelian and Abelian flavor models that incorporate this 
mechanism.
\end{abstract}

\end{titlepage}
\renewcommand{\thepage}{\roman{page}}
\setcounter{page}{2}
\mbox{ }

\vskip 1in

\begin{center}
{\bf Disclaimer}
\end{center}

\vskip .2in

\begin{scriptsize}
\begin{quotation}
This document was prepared as an account of work sponsored by the United
States Government. While this document is believed to contain correct 
information, neither the United States Government nor any agency
thereof, nor The Regents of the University of California, nor any of their
employees, makes any warranty, express or implied, or assumes any legal
liability or responsibility for the accuracy, completeness, or usefulness
of any information, apparatus, product, or process disclosed, or represents
that its use would not infringe privately owned rights.  Reference herein
to any specific commercial products process, or service by its trade name,
trademark, manufacturer, or otherwise, does not necessarily constitute or
imply its endorsement, recommendation, or favoring by the United States
Government or any agency thereof, or The Regents of the University of
California.  The views and opinions of authors expressed herein do not
necessarily state or reflect those of the United States Government or any
agency thereof, or The Regents of the University of California.
\end{quotation}
\end{scriptsize}

\vskip 2in

\begin{center}
\begin{small}
{\it Lawrence Berkeley Laboratory is an equal opportunity employer.}
\end{small}
\end{center}

\newpage
\renewcommand{\thepage}{\arabic{page}}
\setcounter{page}{1}
\section{Introduction} \label{sec:intro} \setcounter{equation}{0}

Symmetry is a powerful tool for understanding the physical world, even
when the symmetry in question is known to be broken.  However, many
candidate fundamental theories are incomplete, or flawed, because we
do not know how their symmetries are broken --- the origin of symmetry
breaking is perhaps the greatest gap in our understanding of nature.

The spontaneous breaking of approximate light-quark flavor symmetries
in QCD, leading to light pions and kaons, is the only case in nature
where we know the underlying theory of symmetry
breaking~\cite{Georgi}. The origin of SU(2)$_L\times$U(1)$_Y$ electroweak
symmetry breaking, leading to the $W$ and $Z$ masses, and of the
U(3)$^5$ flavor symmetry breaking, leading to the quark and lepton
masses, is unknown. There are only a few candidate field theory
mechanisms for such symmetry breakings.  Symmetries are apparently
easily broken by the vacuum expectation values of elementary scalar
fields~\cite{Higgs}, but this alone is unsatisfactory, as it
does not provide an understanding for the mass scale of the associated
symmetry breaking.  Without such information, we do not have an
understanding of the basic mass scales of nature.

The only known way to generate symmetry breaking mass scales in
quantum field theory is by dimensional transmutation, frequently, but
not always, involving strongly interacting dynamics.  Examples of such
dynamical symmetry breaking are provided by QCD, and by theories of
dynamical supersymmetry breaking. In supersymmetric theories, once
soft scalar masses are induced from supersymmetry breaking, gauge and
global symmetries may be broken by having further interactions which
evolve these squared masses negative, thus dynamically generating new
symmetry breaking~\cite{RadiativeBreaking}. For example, the large top
Yukawa coupling has been used to drive the Higgs mass-squared
negative, breaking electroweak symmetry.  Much model building has
centered around this two stage breaking of symmetries: first
supersymmetry is broken to generate the soft squared masses, then
further, non-gauge interactions give radiative corrections so that
the squared masses become negative.

In view of the importance of symmetry breaking, it is striking that
certain strong supersymmetric gauge interactions necessarily force a
direct breaking of symmetries \cite{seiberg}.  This does not require
supersymmetry breaking, nor any other interactions beyond the
supersymmetric gauge interactions.\footnote{In certain other theories,
symmetry breaking can occur by the combination of supersymmetric gauge
interactions and superpotential interactions. These have recently been
used to study the breaking of grand unified symmetries \cite{cheng}.}
For example, in supersymmetric QCD with an equal number of flavors and
colors, the strong gauge interaction forms bound state mesons and
baryons, $T$ and $B$, and induces vevs for some of their scalar
components.  This direct forcing of symmetry breaking offers a new 
avenue for exploring the origins of gauge and flavor symmetry breaking. In 
this paper, we use this strong dynamics to construct realistic theories of
flavor.

In supersymmetric extensions of the standard model, flavor symmetries
are in general broken by squark and slepton mass matrices, $m^2$, as
well as by Yukawa matrices, $h$, which generate the quark and lepton
masses. In this paper we study theories where the form of both $m^2$
and $h$ are governed by some fundamental global flavor symmetry group,
$G_F$, and its breaking pattern. We take the preons, $p$, and the
bound states, $T$ and $B$, of some new strong gauge force to transform
non-trivially under $G_F$. The theory contains the most general set of
interactions which are gauge and $G_F$ invariant, both F and D
terms, including non-renormalizable operators, scaled by inverse
powers of the cutoff $M_*$, which we take to be the reduced Planck
scale $M_{Pl}/\sqrt{8\pi}\simeq 2.4\times 10^{18}$ GeV. In the
fundamental theory, the scalar mass and Yukawa matrices can be written
as field dependent polynomials, $m^2(p/M_*)$ and $h(p/M_*)$, where $p$
is a preon field. At the scale $\Lambda$ of the new strong force,
these matrices become polynomials in the meson and baryon fields,
\begin{equation}
m^2 = m^2 \left({\Lambda T \over M_*^2}, {\Lambda^{N-1} B \over M_*^N}\right)
\label{m^2}
\end{equation}
\begin{equation}
h = h \left({\Lambda T \over M_*^2}, {\Lambda^{N-1} B \over M_*^N}\right),
\label{h}
\end{equation}
where $N$ is the number of preons in a baryon.  The new strong force
constrains $T$ and $B$ to acquire vevs so that these fields become
flavon fields, spontaneously breaking the flavor group $G_F$.
However, there is a large vacuum degeneracy, so that $m^2$ and $h$
become functions on the moduli space. The main phenomenological problem is
to lift this vacuum degeneracy, so that for a certain choice of 
$T, B$ and $G_F$, (\ref{m^2}) and (\ref{h}) give realistic masses.

In the next section we elaborate on the framework for symmetry
breaking and solving the vacuum alignment problem. In Sections 3 and 4
we give explicit realistic theories of flavor, based on non-Abelian
and Abelian $G_F$, respectively. Our vacuum alignment mechanism
results in all non-zero vevs of $T$ and $B$ being of order $\Lambda$.
The flavor group is broken at a single scale -- there is no hierarchy
of symmetry breaking scales -- so that all the small parameters of
$m^2$ and $h$ are derived from $\Lambda/M_*$. For example, a term in
(\ref{m^2}) or (\ref{h}) involving $n_T$ meson fields and $n_B$ baryon 
fields leads to a dimensionless coefficient of size 
$(\Lambda/M_*)^{2n_T + Nn_B}$.  The hierarchy of quark and lepton masses 
arises because of the small value of $\Lambda/M_*$, because mesons and 
baryons contain different numbers of preons, and because the $G_F$ quantum 
number assignments lead to interactions with differing numbers of mesons 
and baryons.

\section{Framework} \label{sec:frame} \setcounter{equation}{0}

In this section we outline our general approach for breaking 
flavor symmetries dynamically in models with composite flavon fields.
We give explicit examples of viable models that incorporate
this mechanism in the following section.

The sector of the theory that is responsible for confinement is a
supersymmetric SU(N) gauge theory with $N$ flavors.  The nonanomalous
global symmetries of the theory are
$G=$SU(N)$_p\times$SU(N)$_{\overline{p}}\times$U(1)$_B\times$U(1)$_R$,
where the first U(1) factor is the analog of baryon number in ordinary
QCD, and the second U(1) is an R-symmetry.  The transformation
properties of the preons and their bound states under the global
symmetries $G$ are shown in Table~\ref{table1}.  Notice that there are
$N^2$ meson fields with zero baryon number, transforming as an $(N,N)$
under the two global SU(N) groups, and a baryon-antibaryon pair that
are singlets under the two SU(N)s.

\begin{table}
\centering
\begin{tabular}{c|c|cccc}
\hline\hline
               & SU(N) & SU(N)$_p$ & SU(N)$_{\overline{p}}$ & 
U(1)$_B$ & U(1)$_R$ \\ \hline
$p$            & $\Box$ &  $\Box$ & $1$ & $1$ & $0$ \\
$\overline{p}$ & $\stackrel{-}{\Box}$ & $1$ & $\Box$ & $-1$ & $0$ \\
\hline
$p\overline{p}$ & $1$ & $\Box$ & $\Box$ & $0$ & $0$ \\
$p^N$           & $1$ & $1$ & $1$ & $N$ & $0$ \\
$\overline{p}^N$ & $1$ & $1$ & $1$ & $-N$ & $0$ \\
\hline\hline
\end{tabular}
\caption{SU(N) with $N$ flavors. \label{table1}} 
\end{table}

This confining theory has two features that are particularly relevant
to model building.  First, an SU(N) gauge theory with $N$ flavors has
no dynamically generated superpotential.  This follows from the fact
that all the preons in Table~\ref{table1} have R-charge $0$, so that
it is not possible to write down an invariant combination of the
fields that have R-charge $2$.  Secondly, the vacuum manifold of the
theory is distorted by quantum mechanical effects so that the origin
of field space is excluded~\cite{seiberg}. Classically, we have the
identity
\begin{equation}
\det (p \overline{p}) - p^N \overline{p}^N = 0
\end{equation}
which we can rewrite in terms of canonically normalized meson and
baryon fields as
\begin{equation}
\det M - \Lambda^{N-2} B \overline{B} = 0 .
\end{equation}
Quantum mechanically, this relation is modified, and becomes
\begin{equation}
\det M - \Lambda^{N-2} B \overline{B} = \Lambda^N  \,\,\, .
\label{eq:qcon}
\end{equation}
Notice that there is no symmetry which prevents the right-hand side
of Eq.~(\ref{eq:qcon}) from becoming nonzero.  Furthermore 
this modified constraint is necessary if we are to properly recover 
the Affleck-Dine-Seiberg superpotential \cite{ads} when we decouple 
one flavor, beginning with the SU(N) theory with equal numbers of 
flavors and colors.

We learn from Eq.~(\ref{eq:qcon}) that some of the meson
and baryon fields acquire vevs, breaking the original global 
symmetry $G$.  If the preons transform nontrivially under a
flavor symmetry group $G_F$, then meson and baryon vevs 
may break the flavor symmetry as well.  If we interpret $G$ as an 
accidental symmetry of the sector responsible for confinement, while 
$G_F$ is respected by all the interactions of the theory,  then some 
of the mesons and baryons may couple to ordinary matter and serve as 
flavon fields.  Yukawa couplings may arise via Planck-suppressed operators, 
as described in Section~1, so that the small parameter that characterizes 
flavor symmetry breaking is the ratio of the confinement scale $\Lambda$ 
to the reduced Planck mass $M_*$.

The ambiguity that must now be resolved is the precise set of composites 
that actually acquire vevs.  For example, Eq.~(\ref{eq:qcon}) is satisfied
by a point in field space where the baryons $B$ and $\overline{B}$ acquire 
confinement-scale vevs, while the mesons remain at the origin. This vacuum 
would not be particularly useful if we were to construct a model in 
which only the mesons coupled to ordinary matter.  In a viable model, we 
must sufficiently reduce this vacuum degeneracy so that the flavor symmetry 
breaking fields which couple to ordinary matter are forced to get vevs.  The 
models that we present in the next section achieve this in two steps:

First, we introduce additional fields $X_{j}$ $(j=0,1,2,\ldots)$, that 
couple to the preons via nonrenormalizable superpotential interactions.  
Since the preon fields have U(1)$_R$ charge $0$, we will take the
fields $X$ to have $R$-charge $2$.  We impose the U(1)$_R$ symmetry
so that all of the preonic operators involve one of the
$X$ fields\footnote{We assume that this symmetry is spontaneously
broken in the hidden sector, so that we generate gaugino masses,
trilinear scalar interactions, etc.}.  We will assume that $X_0$ is a 
singlet under the non-$R$ symmetries shown in Table~\ref{table1}, while 
the remaining $X_j$ transform nontrivially under $G_F$.  The $X$
fields will be responsible for restricting the moduli space such
that the desired set of mesons and baryons develop vacuum expectation
values when the scalar potential is minimized.

The F-flatness conditions for the $X$ fields significantly
reduce the original supersymmetric vacuum degeneracy.  Consider the 
superpotential interactions for the field $X_0$.  In the models
of interest, these will be of the form
\begin{equation}
W_0 = \left(\frac{1}{M_*}\right)^{2N-2}
\left(\frac{\Lambda}{M_*}\right)^N
\left[\sum_j G_j + \Lambda^{N-2} B \overline{B} \right] X_0 \,\,\, 
\label{eq:w0}
\end{equation}
where the $G_i$ represent all possible flavor-group invariant combinations 
of the meson fields involving $2N$ preons.  In the models we will consider, 
these interactions will be the ones of lowest order in $1/M_*$ that are 
allowed by the flavor symmetry.  Other interactions, such as direct 
$G_F$-invariant couplings between the baryons and mesons, will arise at 
higher order, and will be suppressed.  Note that we have omitted a 
Planck-scale linear term for $X_0$, which can be forbidden by imposing an 
anomalous discrete symmetry, as we will see explicitly in the next section.  
Notice that the F-flatness condition for $X_0$ together with the
quantum-modified constraint (\ref{eq:qcon}) yield two restrictions on
the set of invariants $G_i$, $B\overline{B}$.  Thus, we have succeeded
in reducing the vacuum degeneracy by one degree of freedom. The $X_0$ field
orients the vacuum so that at least some of the mesons have non-vanishing 
vevs.

Now we introduce additional fields $X_j$, that transform
nontrivially under the flavor group $G_F$.  These lead to
additional superpotential couplings of the form
\begin{equation}
W_j = X_j \sum_i G'_i
\end{equation}
where the $G'_i$ represent all possible baryon and meson interactions
with the appropriate quantum numbers to couple to $X_j$.  We have absorbed 
powers of $\Lambda$ and $M_*$ into the definition of the $G'_i$ for 
notational convenience.  In a successful model, we introduce enough 
nontrivial constraints in this way such that the flavor invariant 
combinations of the mesons and baryons shown in brackets in 
Eq.~(\ref{eq:w0}) acquire vevs {\em individually},
\begin{eqnarray}
& B\overline{B} \sim \Lambda^2 & \nonumber \\
& G_1 \sim \Lambda^N & \nonumber \\
& G_2 \sim \Lambda^N & \nonumber \\
& \mbox{etc.} &      ,
\label{eq:morevevs}
\end{eqnarray}
while all the $X$ field vevs vanish.  This result should remain
valid provided that the K\"ahler metric is positive definite in
the region of field space where we have located the minimum.
Since the K\"ahler potential is not calculable for field amplitudes 
of the same order as the confinement scale, we take this positivity 
requirement as a mild assumption.  Note also that if too many
F-flatness constraints are added, it is possible that the resulting
superpotential may break supersymmetry.  This would lead to direct,
flavor-dependent couplings of fields with large F components to
ordinary matter, which would not be desirable.  In all the models we
consider, supersymmetry will remain unbroken after the effects of the
$X$ fields are taken into account.

Once we have arranged for each gauge invariant combination of
the mesons and baryons to acquire vevs, we must lift the remaining vacuum 
degeneracy.  Notice that given any point in the moduli space defined 
by Eq.~(\ref{eq:morevevs}), we can reach another point by transforming the 
fields under the complexification of the flavor group.  If we include 
positive soft supersymmetry-breaking squared masses for the 
the composite flavon fields, the complexified symmetry will 
be broken, and this last flat direction will be lifted. (We justify this
procedure below.)  To make this point concrete, imagine we have a theory 
with three flavons, $\phi_0$, $\phi_+$ and $\phi_-$, where the subscript 
indicates the charge under some U(1) symmetry.  Now assume that the moduli 
space is constrained such that $\phi_0^3 \sim \Lambda^3$ and 
$\phi_0\,\phi_+\phi_- \sim \Lambda^3$.  The remaining flat direction 
corresponds to a rescaling of $\phi_+$ and $\phi_-$, which is generated 
by the complexification of the U(1) symmetry. The complexified symmetry 
is explicitly broken by the soft mass terms 
$V_{soft} = m^2_{soft} (|\phi_0|^2 + |\phi_+|^2+|\phi_-|^2)$, and 
minimization of the full potential then yields 
$\phi_+ \sim \phi_- \sim \phi_0 \sim \Lambda$, as desired.  

This last step may be questioned since the form of the soft supersymmetry 
breaking interactions in the confining theory are not determined by any
symmetry argument.  However, we may justify 
our qualitative result by considering the behavior of the theory 
in the limit of large field amplitudes.  Our constraints on the 
gauge-invariant products of the fields $G_i \sim \Lambda^N$ imply 
that varying any moduli field away from $\Lambda$ forces some field 
to acquire a vev greater than $\Lambda$.   In the limit of large
field amplitudes, this corresponds to at least some of the preons $p$ 
acquiring large expectation values as well, $p > \Lambda$.   In the same 
limit we expect there will be soft supersymmetry breaking squared masses 
for the preon fields, so $V_{soft} \sim m^2_{soft} |p|^2$.   Again assuming 
positive $m^2_{soft}$, the potential grows as we take any $p$ larger
than $\Lambda$, and we conclude that our previous result is energetically 
favored.  This conclusion is consistent with the assumption made in 
Ref.~\cite{peskin} that minimization of a potential that includes soft 
supersymmetry breaking masses for the composite fields should 
lead to correct qualitative results, even when field amplitudes are 
of the same order as the confinement scale.  Therefore, for the purpose 
of calculation, we will assume soft masses for the composite fields, but 
the reader should keep in mind that the results are supported by 
this more general argument.

Finally, we will make the simplifying assumption that trilinear scalar 
interactions (A-terms) can be neglected in the potential. Any  minimum 
of the potential that we find in the absence of A-terms will remain 
at least a local minimum for small but nonvanishing A parameters.
This will be sufficient for our purposes.  We will not attempt to find 
the explicit conditions implied by vacuum stability on the possible 
trilinear scalar interactions in the flavor sector when the
$A$ parameters are large.

After taking into account both the F-flatness conditions for the
$X$ fields, and the effect of soft supersymmetry-breaking scalar
masses, it is often the case that the desired composite fields
will each be forced to acquire a vev of order $\Lambda$.  We will now 
present two complete models that successfully incorporate the flavor 
symmetry breaking mechanism described in this section.

\section{Non-Abelian Model} \label{sec:model_su(3)} 
\setcounter{equation}{0}

The models we present in this and the next section are based on
SU(3) supersymmetric QCD with three flavors.  The global symmetries 
of the strong interaction are
\begin{equation}
G = {\rm SU(3)}_p\times {\rm SU(3)}_{\overline{p}}\times {\rm U(1)}_B
\times {\rm U(1)}_R 
\end{equation}
and
\begin{equation}
G_A = {\rm U(1)}_A  \,\,\, ,
\end{equation}
where $G_A$ is the anomalous U(1) symmetry corresponding to axial
phase rotations on $p$ and $\overline{p}$.  In each of the models
we present, the action of the flavor group $G_F$ on the preon fields 
will be isomorphic to a subgroup of $G \times G_A$.  However,
one should keep in mind that the ordinary fermions will transform under 
$G_F$ even though they do not transform under the global symmetries of 
the confining flavor sector.

\begin{table}
\centering
\begin{tabular}{l| l l}
\multicolumn{3}{c}{MSSM Fields} \\ \hline \hline
{} & {${\rm G_{SM}}$} & {${\rm G_F}$} 
 \\ \hline 
  $Q^i$ & 
  $(\Box, \Box, \frac{1}{6})$ & 
  $(\stackrel{-}{\Box},-1,2,-)$
 \\
  $Q^3$ &
  $(\Box, \Box , \frac{1}{6})$ &
  $({\bf 1},0,0,+)$
 \\
  $U^i$ &
  $(\stackrel{-}{\Box}, {\bf 1}, -\frac{2}{3})$ &
  $(\stackrel{-}{\Box},-1,2,-)$
 \\
  $U^3$ &
  $(\stackrel{-}{\Box}, {\bf 1}, -\frac{2}{3})$ &
  $({\bf 1},0,0,+)$
 \\
  $D^i$ &
  $(\stackrel{-}{\Box}, {\bf 1}, \frac{1}{3})$ &
  $(\stackrel{-}{\Box},-1,2,-)$
 \\
  $D^3$ &
  $(\stackrel{-}{\Box}, {\bf 1}, \frac{1}{3})$ &
  $({\bf 1},2,2,-)$
 \\ \hline
  \multicolumn{3}{c}{} \\
\multicolumn{3}{c}{Composite Fields} \\ \hline \hline
{} & {${\rm G_{SM}}$} & {${\rm G_F}$} 
 \\ \hline 
  $\phi_i$ & $({\bf 1}, {\bf 1}, 0)$ &
  $(\Box,1,-2,-)$
 \\
  $\tilde{\phi}_i$ & $({\bf 1}, {\bf 1}, 0)$ &
  $(\Box,-2,1,-)$
 \\
  $A$ & $({\bf 1}, {\bf 1}, 0)$ &
  $({\bf 1},1,1,-)$
 \\
  $S_{ij}$ & $({\bf 1}, {\bf 1}, 0)$ &
  $(\Box\hspace{-0.04in}\Box,1,1,-)$
 \\
  $\sigma$ & $({\bf 1}, {\bf 1}, 0)$ &
  $({\bf 1},-2,-2,-)$
 \\
  $B$ & $({\bf 1}, {\bf 1}, 0)$ &
  $({\bf 1},1,-2,-)$
 \\
  $\bar{B}$ & $({\bf 1}, {\bf 1}, 0)$ &
  $({\bf 1},-1,2,+)$
 \\ \hline\hline 
\end{tabular} 
\caption{The transformation properties of the quarks and the
composite states under the standard model gauge group 
$G_{{\rm SM}}=$SU(3)$_C\times$SU(2)$_L\times$U(1)$_Y$ and the flavor 
symmetry $G_F=$SU(2)$_F\times$U(1)$_F\times$U(1)$_{F'}\times Z_2$. Here,
$i=1,2$ is the ${\rm SU(2)_F}$ index. Note that the Higgs fields
$H_1$ and $H_2$ are invariant under ${\rm G_F}$.\label{table:tp}}
\end{table}

The flavor group of the first model is 
\begin{equation}
G_F= {\rm SU(2)}_F\times {\rm U(1)}_F\times {\rm U(1)}_{F'}\times Z_2
\,\,\, .
\end{equation}
The transformation properties of the MSSM superfields (as well as those of 
the composite states discussed later) are shown in Table~\ref{table:tp}.  The 
lighter two generations of the matter fields  ($Q^i$, $U^i$, and $D^i$, 
with $i=1,2$) transform as doublets under SU(2)$_F$, while the third 
generation fields ($Q^3$, $U^3$, and $D^3$) are singlets.  The
${\bf 2}+{\bf 1}$ representation structure provides a natural degeneracy 
between squark masses of the first and second generations in the flavor 
symmetric limit~\cite{dlk}.  This leads to a suppression of flavor changing
neutral current effects when the flavor symmetries are broken.
The remaining group factors, U(1)$_F\times$U(1)$_{F'}$, are used 
to obtain realistic Yukawa textures.  The fields $Q^i$, $U^i$, $D^i$, and 
$D^3$ transform nontrivially under the two flavor U(1) factors, while 
$Q^3$, $U^3$ and the ordinary Higgs fields are $G_F$ invariant.  The top
quark Yukawa coupling is invariant under the flavor symmetry, and hence
can be of order one, while the other Yukawa elements will be 
suppressed by the ratios of flavon vevs to $M_*$.

If we consider the preonic sector alone, the flavor symmetry can be 
identified with a subgroup of $G\times G_A$.  We first decompose each
SU(3) factor into its SU(2)$\times$U(1) subgroup:
\[
{\rm SU(3)}_p\times{\rm SU(3)}_{\overline{p}} \times {\rm U(1)}_B
\rightarrow \left[ {\rm SU(2)}\times{\rm U(1)}\right]_p \times
\left[ {\rm SU(2)}\times{\rm U(1)}\right]_{\overline{p}} \times
{\rm U(1)}_B
\]
The flavor SU(2) is simply the diagonal subgroup of 
SU(2)$_p\times$SU(2)$_{\overline{p}}$.  The two flavor U(1) factors are 
different linear combinations of U(1)$_B$, U(1)$_p$ and U(1)$_{\overline{p}}$. 
The charges under the flavor U(1)s are defined by
\begin{equation}
Q_{\rm F} = 2\sqrt{3}Q_p + \frac{1}{3}Q_{\rm B}  
\,\,\,\,\, \mbox{ and } \,\,\,\,\,  
Q_{\rm F'} = 2\sqrt{3}Q_{\bar{p}} - \frac{2}{3}Q_{\rm B} \,\,\, ,
\end{equation}
where $Q_p$ and $Q_{\overline{p}}$ are the eigenvalues of the
$T^8$ generators of SU(3)$_p$ and SU(3)$_{\overline{p}}$, respectively.

The quantum numbers of the preons $p$ and $\overline{p}$ under 
SU(3)$\times G_F$ are given by
\begin{eqnarray}
p^i &\sim & (\Box,\Box,\frac{4}{3},-\frac{2}{3},-) \\ 
p   &\sim & (\Box,{\bf 1},-\frac{5}{3},-\frac{2}{3},-) \\
\bar{p}^i & \sim & (\stackrel{-}{\Box},\Box,-\frac{1}{3},\frac{5}{3},+) \\
\bar{p} & \sim & (\stackrel{-}{\Box},{\bf 1},-\frac{1}{3},\frac{4}{3},+).
\end{eqnarray}
Notice that the $Z_2$ factor is a symmetry under which all the preons
are odd and all anti-preons are even; this is a discrete
subgroup of U(1)$_A\times$U(1)$_B$.  Once the SU(3) gauge 
group becomes strong at the scale $\Lambda$, the preons form composite 
states: 
 \begin{eqnarray}
  && S_{ij}\sim \Lambda^{-1} (p_{i}\bar{p}_{j}+p_{j}\bar{p}_{i}),
 \label{S_ij} \\ &&
  A\sim \Lambda^{-1} \epsilon^{ij}(p_{i}\bar{p}_{j}),
 \\ &&
  \phi_i\sim\Lambda^{-1} (p_i\bar{p}),
 \\ &&
  \tilde{\phi}_i\sim \Lambda^{-1} (p\bar{p}_i),
 \\ &&
  \sigma \sim \Lambda^{-1} (p\bar{p}),
 \\ &&
  B \sim \Lambda^{-2} \epsilon^{ij} (p_i p_j p),
 \\ &&
  \bar{B} \sim \Lambda^{-2} \epsilon^{ij} (\bar{p}_i \bar{p}_j \bar{p}),
 \label{B-bar}
 \end{eqnarray}
where $i$ and $j$ are SU(2)$_F$ flavor indices.  The composite
fields have been given canonical mass dimension by including
appropriate powers of $\Lambda^{-1}$.  The transformation properties 
of the composite states under the flavor symmetry are also summarized
in Table~\ref{table:tp}.

Given these quantum number assignments, and our assumption that
Planck-scale physics induces all operators that are consistent with
the symmetries, some of the composite states above can serve
as flavon fields.  The $G_F$-allowed couplings that can contribute to 
the Yukawa matrices are summarized as follows:
\begin{equation}
h_u \sim \left(
\begin{array}{cc|c} 
0   & B^2 & 0  \\
- B^2 & \phi_2\phi_2 & \phi_2 \\
\hline
0 & \phi_2 & 1 \end{array}\right)
\label{eq:hu}
\end{equation}
\begin{equation}
h_d \sim \left(
\begin{array}{cc|c} 
0   & B^2 & 0  \\
-B^2 & \phi_2\phi_2 & \phi_2 \sigma \\
\hline
0 & \phi_2 & \sigma \end{array}\right)  \,\,\, .
\label{eq:hd}
\end{equation}
We have not shown couplings to $\phi_1$, since we will
always work in a basis where the $\phi_1$ vev vanishes.
We have also temporarily suppressed the factors of $\Lambda$ and $M_*$ 
in each entry, which depend on the dimensionality of the original
preonic interaction.  Note that if a composite field above acquires a
vev of order $\Lambda$, then the size of the corresponding
Yukawa entry will be $(\Lambda/M_*)^n$, where $n$ is the total
number of preons involved in the preonic higher-dimension operator.  
To obtain a realistic theory we need only lift the vacuum degeneracy such
that $B, \phi$ and $\sigma$ are all forced to acquire vevs of order
$\Lambda$. There may be many ways to accomplish this; below we provide an
explicit example.

Since the confining sector of our model is of the type described in 
Section~\ref{sec:frame}, the moduli space of the composite states is 
restricted by a quantum-modified constraint \cite{seiberg}.
The important point is that the origin of field space is excluded, so 
that the flavor symmetries are guaranteed to break.  The constraint 
is realized by a dynamically generated Lagrange multiplier term in 
the superpotential
 \begin{eqnarray}
  W_{\rm dyn}&=& \eta \left[ C_{M^3}
    ( \epsilon^{ij}A\phi_i\tilde{\phi}_j
    + A^2 \sigma
    + \epsilon^{ij}\epsilon^{kl}
      S_{ik}\phi_j\tilde{\phi}_l \right.
  \nonumber \\ &&
\left. + \epsilon^{ij}\epsilon^{kl}S_{ik}S_{jl}\sigma)    
+ C_{B\bar{B}} \Lambda B\bar{B}
    - \Lambda^3 \right],
 \label{W_dyn}
 \end{eqnarray}
where $\eta$ is the Lagrange multiplier field, and $C$'s are ${\cal
O}(1)$ coefficients that arise from the dynamics of confinement.

As we described in Section~\ref{sec:frame}, the constraint equation
alone leaves us with a rather large vacuum degeneracy, and the
possibility that we will not obtain a viable pattern of flavor
symmetry breaking.  We will remove most of these flat directions
by introducing several $X$ fields, to place additional 
constraints on the composite states.  Perhaps the simplest
set of $X$ fields is given by
\begin{eqnarray}
X_0 & \sim & ({\bf 1},0,0,-) \nonumber \\
X_1 & \sim & ({\bf 1},1,1,+) \nonumber \\
X_2 & \sim & (\Box\hspace{-0.04in}\Box,-1,-1,-)  \,\,\, ,
\end{eqnarray}
where we have shown the transformation properties under 
${\rm G_F}$ in parentheses.  We can now write down the
following interaction terms for the preons:
\begin{eqnarray}
  W_0 &=& \frac{1}{M_*^4} X_0
    [ (\epsilon^{ij}p_{i}\bar{p}_{j}) 
    \epsilon^{kl} (p_k\bar{p})(p\bar{p}_l)
    + (\epsilon^{ij}p_{i}\bar{p}_{j})^2 (p\bar{p})
    + \epsilon^{ij}\epsilon^{kl} (p_{i}\bar{p}_{k}+p_{k}\bar{p}_{i})
    (p_j\bar{p}) (p\bar{p}_l) \label{eq:w0eq}
 \nonumber \\ &&
    + \epsilon^{ik}\epsilon^{jl}
    (p_{i}\bar{p}_{j}+p_{j}\bar{p}_{i}) 
    (p_{k}\bar{p}_{l}+p_{l}\bar{p}_{k}) (p\bar{p})
    + (\epsilon^{ij} p_i p_j p) 
    (\epsilon^{kl} \bar{p}_k \bar{p}_l \bar{p}) ]
       \\ 
W_1 & = &
    \frac{1}{M_*^2} X_1
    [(\epsilon^{ij}p_{i}\bar{p}_{j}) (p\bar{p})
    + \epsilon^{ij} (p_i\bar{p})(p\bar{p}_j)] \\ 
W_2 & = &
    \epsilon^{ik}\epsilon^{jl}
    X_{2,ij} (p_{k}\bar{p}_{l}+p_{l}\bar{p}_{k}),
\end{eqnarray}
where we have omitted unknown ${\cal O}(1)$ coefficients. Notice that
there is no linear term in $X_0$ due to the discrete $Z_2$ symmetry.
Without this symmetry, the interaction $M_*^2X_0$ would also be
allowed, and the argument presented below would break down.  Note
that the $Z_2$ symmetry has no significant effect on the mass matrix
textures that we obtain in either of our models.

After confinement, F-flatness conditions for the Lagrange
multiplier field $\eta$ and the fields $X_j$  give us the 
following four equations of motion for composite states:
 \begin{eqnarray}
 &&
  C_{M^3} \epsilon^{ij}A\phi_i\tilde{\phi}_j
    + C_{M^3} A^2 \sigma + C_{B\bar{B}} \Lambda B\bar{B}
    - \Lambda^3 = 0,
 \label{F_X0=0} \\ &&
  C'_{A\phi\tilde{\phi}} \epsilon^{ij}A\phi_i\tilde{\phi}_j
    + C'_{A^2\sigma} A^2 \sigma + C'_{B\bar{B}} \Lambda B\bar{B}
     - \Lambda^3 = 0,
 \label{F_X1=0} \\ &&
  C''_{\phi\tilde{\phi}} \epsilon^{ij}\phi_i\tilde{\phi}_j
  + C''_{A\sigma} A \sigma = 0,
 \label{F_X2=0} \\ &&
  S_{ij} = 0.
 \label{F_X3=0}
 \end{eqnarray}
Here the $C'$ and $C''$ are also ${\cal O}(1)$ coefficients\footnote{
In writing down the low energy description of the operator 
$X_0[ppp\bar{p}\bar{p}\bar{p}]/M^4_*$, we have included a linear 
term $X_0 \Lambda^6/M_*^4$.   This term can be justified by treating 
$\Lambda$ as a spurion under the anomalous axial U(1) symmetry of the 
dynamical sector, and including the most general set of
invariant interactions.  However, nothing in our analysis changes if such 
a term is absent.}.  Note that we have dropped the terms which depend 
on $S_{ij}$ in Eqs.(\ref{F_X0=0}) -- (\ref{F_X2=0}) by using 
eq.(\ref{F_X3=0}).  We can easily solve eqs.(\ref{F_X0=0}) -- (\ref{F_X2=0}), 
and we obtain 
 \begin{eqnarray}
  && A^2 \sigma \sim \Lambda^3,
 \label{A^2*sigma} \\
  && \epsilon^{ij}A\phi_i\tilde{\phi}_j \sim \Lambda^3,
 \label{a*phi*phi} \\
  && B\bar{B} \sim \Lambda^2,
 \label{BB}
 \end{eqnarray}
neglecting the possibility of accidental cancellations.
 
At this stage, the remaining flat directions correspond to
rescaling of the composite fields, since only their products
are constrained by  Eqs.~(\ref{A^2*sigma}) -- (\ref{BB}).
One might suspect that these flat directions can be lifted by 
including yet higher order corrections to the superpotential.
However, this can never be the case because the remaining flat
directions are protected by a symmetry that is respected by all
F-term contributions to the potential in the supersymmetric limit.  Since 
our model has an SU(2)$_F\times$U(1)$_F\times$U(1)$_{F'}$ global symmetry, 
and the superpotential is holomorphic in the fields, the actual symmetry of 
the superpotential before supersymmetry breaking is the complexification of 
the flavor group.  Since SU(2)$\times$U(1)$\times$U(1) has $5$ generators, 
this symmetry corresponds to 5 complex degrees of freedom in the moduli 
space.  We began with $9$ meson and $2$ baryon fields, and imposed 
$6$ F-flatness conditions (for the fields $\eta$, $X_0$, $X_1$, and the 
three components of $X_2$) leaving $5$ complex degrees of freedom.  Thus, 
we have lifted all the flat directions that are not protected by the 
complexified symmetry.

To lift the flat directions defined by Eqs.~(\ref{A^2*sigma}) -- (\ref{BB}),
we include the soft supersymmetry-breaking scalar masses for the composite 
states, 
\begin{eqnarray}
  V_{\rm soft} = m_{\rm soft}^2
    (|A|^2 + |S|^2 + |\phi|^2 + |\tilde{\phi}|^2 + |\sigma|^2
    + |B|^2 + |\bar{B}|^2)
\label{eq:smasses}
\end{eqnarray}
with $m^2_{{\rm soft}} > 0$\footnote{Here, 
we do not assume universal soft supersymmetry-breaking masses.  The 
$m^2_{{\rm soft}}$ in Eq.~(\ref{eq:smasses}) is understood to be different 
for each of the terms shown.}.  We now minimize the potential above subject 
to the constraints (\ref{A^2*sigma}) -- (\ref{BB}).  The qualitative result 
is easy to understand.  We first use the  ${\rm SU(2)_F}$ symmetry
to work in the basis where $\phi_1=0$. In this basis,
$\tilde{\phi}_2$ appears in Eq.~(\ref{eq:smasses}), but not
in any of the constraints, and is therefore driven to zero.
The non-vanishing elements ($\phi_2$, $\tilde{\phi}_1$, $A$, $\sigma$, $B$, 
and $\bar{B}$) must all be of ${\cal O}(\Lambda)$, so that the constraints 
(\ref{A^2*sigma}) -- (\ref{BB}) are satisfied and $V_{\rm soft}\sim m_{\rm
soft}^2\Lambda^2$.  If any of the fields were to have a vev smaller than
$\Lambda$, the constraint equations assure that another composite have 
a vev larger than $\Lambda$, and we would obtain $V_{\rm soft}> m_{\rm
soft}^2\Lambda^2$.  Therefore, up to an ${\rm SU(2)_F}$ rotation, the
minimum is at
 \begin{eqnarray}
  \phi\sim
  \left( \begin{array}{c} 0 \\ \Lambda \end{array} \right),~~~
  \tilde{\phi}\sim
  \left( \begin{array}{c} \Lambda \\ 0 \end{array} \right),~~~
  A \sim \sigma \sim B \sim \bar{B} \sim \Lambda,~~~
  S_{ij}=0.
\label{eq:themin}
 \end{eqnarray}
This result can be verified by explicit minimization of the
potential, taking into account all the order one parameters.
However, the estimate in Eq.~(\ref{eq:themin}) will be 
sufficient for our purposes.  

The vevs in Eq.~(\ref{eq:themin}) are exactly what we require to
obtain viable textures from Eqs.~(\ref{eq:hu}) and (\ref{eq:hd}).
If we fix the ratio
\begin{equation}
\lambda\equiv\frac{\Lambda}{M_*}<1,
\end{equation}
we obtain
\begin{eqnarray}
  h_u \sim
    \left( \begin{array}{ccc}
    0 & \lambda^6 & 0 \\
    \lambda^6 & \lambda^4 & \lambda^2 \\
    0 & \lambda^2 & 1 \\
    \end{array} \right),~~~
  h_d \sim
    \left( \begin{array}{ccc}
    0 & \lambda^6 & 0 \\
    \lambda^6 & \lambda^4 & \lambda^4 \\
    0 & \lambda^2 & \lambda^2 \\
    \end{array} \right).
 \label{Yud}
 \end{eqnarray}
All the elements in the Yukawa matrices are predicted in terms of
one small parameter $\lambda$, up to unknown coefficients of
order one.  As we will see below, Eq.~(\ref{Yud}) results in 
a realistic pattern of the quark masses and mixing angles,
if $\lambda\sim 0.2-0.3$, and the unknown ${\cal O}(1)$ coefficients 
are chosen appropriately.

We will now consider the pattern of quark masses and mixing angles
more carefully, beginning with the up sector.  The largest element
in the up quark Yukawa matrix is the (3,3) entry, which is of order
$1$, while the other elements are suppressed by powers of $\lambda$.
Thus, $h_u$ has an eigenvalue close to one which we can identify
with the top quark Yukawa coupling.  Next, we consider the 2-3 block,
since the remaining elements are much more suppressed.  The determinant
of this block is of ${\cal O}(\lambda^4)$ indicating there is an
eigenvalue of the same order, which we identify as the charm
quark Yukawa coupling.  The rotation angle involved in the 
diagonalization of this block is ${\cal O}(\lambda^2)$, and
hence $V_{cb}\sim \lambda^2$, if the up sector gives the
dominant contribution.  Finally, we see that $\det h_u \sim
{\cal O}(\lambda^{12})$, which implies that the smallest eigenvalue 
is ${\cal O}(\lambda^8)$.  We identify this with the up quark
Yukawa coupling.  Thus, in our model we find
\begin{equation}
m_u : m_c : m_t \sim \lambda^8 : \lambda^4 : 1
\,\,\,\,\, \mbox{ and } \,\,\,\,\,
V_{cb} \sim \lambda^2.
\end{equation}
As far as the mass eigenvalues and $V_{cb}$ are concerned, the result
for $h_u$ in our model works fairly well.  The only problem is that
the mixing between first and second generations, i.e. the Cabibbo
angle, is ${\cal O}(\lambda^2)$, which is too small if $\lambda\sim
0.2-0.3$.  ($V_{ub}$ on the other hand would be ${\cal O}(\lambda^4)$,
which is acceptable.) Thus, the Cabibbo angle should have its origin
in the down sector. We will come back to this point later.

We may now analyze $h_d$ in the same way.  From the 2-3 block, we
obtain the two larger eigenvalues of $h_d$, which are ${\cal
O}(\lambda^2)$ and ${\cal O}(\lambda^4)$. We identify these as the
Yukawa coupling of the bottom and strange quarks, respectively. Notice
that, with the choice of $\tan\beta\sim 2$, we obtain the correct
value of the ratio of $m_b/m_t$. The 2-3 mixing angle is again ${\cal
O}(\lambda^2)$, and is consistent with the value of $V_{cb}$ for
$\lambda\sim 0.2-0.3$. Finally, we must evaluate the down quark Yukawa
coupling as well as the 1-2 mixing.  Our results in Eq.~(\ref{Yud})
imply naively that the down quark Yukawa coupling is of ${\cal
O}(\lambda^8)$ and the the 1-2 mixing angle of ${\cal O}(\lambda^2)$,
both of which are too small to be consistent with observation. To fix
this problem, we must take into account the possible fluctuations of
the unknown order one coefficients.  If we allow the couplings giving
the (1,2) and (2,1) elements of Eq.~(\ref{eq:hd}) to be enhanced by a
factor of $1/\sqrt{\lambda}\sim 2$, and the (2,2) and (3,2) elements to 
be suppressed by the same amount, we will obtain a Cabibbo angle of
${\cal O}(\lambda)$, and a down quark coupling ${\cal}(\lambda^6
\sqrt{\lambda})$.  Note that the predicted ratio $m_d/m_b \sim
\lambda^4 \sqrt{\lambda}$ is consistent with recent lattice estimates
of the down quark mass~\cite{lattice}.

With this choice for the ${\cal O}(1)$ coefficients, the Yukawa matrix
elements are given more accurately by
\begin{eqnarray}
  h_u \sim
    \left( \begin{array}{ccc}
    0 & \lambda^6 & 0 \\
    \lambda^6 & \lambda^4 & \lambda^2 \\
    0 & \lambda^2 & 1 \\
    \end{array} \right),~~~
  h_d \sim
    \left( \begin{array}{ccc}
    0 & \lambda^{11/2} & 0 \\
    \lambda^{11/2} & \lambda^{9/2} & \lambda^4 \\
    0 & \lambda^{5/2} & \lambda^2 \\
    \end{array} \right).
 \label{Yukawa_with_O(1)}
\end{eqnarray}
We will diagonalize the results shown in Eq.~(\ref{Yukawa_with_O(1)})
when we need to evaluate a squark mass matrix in the quark mass 
eigenstate basis.

Finally, we present the textures for the squark mass matrices.
The soft supersymmetry-breaking masses originate from D-terms
interactions, which are not required to be holomorphic 
functions of the flavon fields.  For example, the leading
contributions to the left-handed squark masses are given
by the operators
\begin{eqnarray}
  V_{\tilde{Q}\tilde{Q}^*} &\sim& 
    \tilde{m}^2 \Bigg[ c_0 |\tilde{Q}^1|^2 + c_0 |\tilde{Q}^2|^2 
    + c_3 |\tilde{Q}^3|^2
 \nonumber \\ &&
    + \frac{\Lambda^2}{M_*^4}
    (\phi_i\tilde{Q}^i) (\phi_j\tilde{Q}^j)^*
    + \frac{\Lambda^2}{M_*^4}
    (\tilde{\phi}_i\tilde{Q}^i) (\tilde{\phi}_j\tilde{Q}^j)^*
 \nonumber \\ &&
    + \frac{\Lambda^2}{M_*^4}
    (\epsilon_{ij}\phi^{*i}\tilde{Q}^j)
    (\epsilon_{kl}\phi^{*k}\tilde{Q}^l)^*
    + \frac{\Lambda^2}{M_*^4}
    (\epsilon_{ij}\tilde{\phi}^{*i}\tilde{Q}^j)
    (\epsilon_{kl}\tilde{\phi}^{*k}\tilde{Q}^l)^*
 \nonumber \\ &&
    + \frac{\Lambda^6}{M_*^{10}}
    \left\{
    (\phi_i\tilde{Q}^i) (\epsilon_{jk}\phi^{*j}\tilde{Q}^k)^* B^{*2}
    + h.c. \right\}
 \nonumber \\ &&
    + \frac{\Lambda^5}{M_*^8}
    \left\{
    (\epsilon_{ij}\phi^{*j}\tilde{Q}^k) \tilde{Q}^{3*}
    B \overline{B}^*
    + h.c. \right\}
 \nonumber \\ &&
    + \frac{\Lambda^2}{M_*^4}
    \left\{ (\tilde{\phi}_i\tilde{Q}^i) \tilde{Q}^{3*} + h.c. \right\}
    \Bigg] \,\,\, ,
 \end{eqnarray}
where $\tilde{m}$ is the typical scale of the squark masses.  We have
only shown order one coefficients explicitly in the flavor-invariant
terms ($c_0$ and $c_3$) to remind the reader that the first two
generation scalars are degenerate in the flavor symmetric limit, 
while the third generation scalar is unconstrained.  After
flavor symmetry breaking, the operators above lead to the
texture
\begin{eqnarray}
  (\tilde{M}_q^2)^0_{LL} \sim
   \tilde{m}^2 \left( \begin{array}{ccc}
    c_0 + \lambda^4 & \lambda^{10} & \lambda^{8} \\
    \lambda^{10} & c_0 + \lambda^4 & \lambda^2 \\
    \lambda^{8} & \lambda^2 & c_3
   \end{array}\right),
 \end{eqnarray}
where the powers of $\lambda$ indicate the correction to the flavor 
invariant result with ${\cal O}(1)$ coefficients suppressed.  It is
straightforward to repeat this analysis for the right-handed squarks,
and we obtain
\begin{eqnarray}
  (\tilde{M}_u^2)^0_{RR}  \sim
   \tilde{m}^2 \left( \begin{array}{ccc}
    c'_0 + \lambda^4 & \lambda^{10} & \lambda^{8} \\
    \lambda^{10} & c'_0 + \lambda^4 & \lambda^2 \\
    \lambda^{8} & \lambda^2 & c'_3
   \end{array}\right)
\end{eqnarray}
and
\begin{eqnarray}
  (\tilde{M}_d^2)^0_{RR} \sim 
   \tilde{m}^2 \left( \begin{array}{ccc}
    c''_0 + \lambda^4 & \lambda^{10} & \lambda^{10} \\
    \lambda^{10} & c''_0 + \lambda^4 & \lambda^4 \\
    \lambda^{10}  & \lambda^4 & c''_3
   \end{array}\right).
\end{eqnarray}

We may now consider the bounds from flavor changing neutral
current processes.  We define the parameters
\begin{eqnarray}
 && 
  (\delta_{ij}^q)_{XX} \equiv
   |(\tilde{M}_q^2)_{XX,ij}| / \tilde{m}^2~~~(X=L,R),
 \\ &&
  \overline{\delta}_{ij}^q \equiv
   \{(\delta_{ij}^q)_{LL} (\delta_{ij}^q)_{RR}\}^{1/2},
 \end{eqnarray}
where $q=u,d$. Note that the absence of the superscript $0$ above $\tilde{M}$
indicates that the scalar mass matrices are to be evaluated in the {\em quark
mass eigenstate basis}.  The $\delta$ parameters corresponding to
1-2 and 1-3 scalar mass matrix elements are constrained by neutral 
pseudoscalar meson mixing to be less than $10^{-1}$ -- $10^{-3}$, depending 
on the superparticle mass spectrum.  Typical upper bounds are given in
Table~\ref{table:delta}.  

We see that the off-diagonal elements are small enough to satisfy the
experimental constraints, with the parameter $\zeta \equiv c''_3/c''_0
=1.3$.  This ratio is not constrained by the flavor symmetry, and must
be mildly adjusted (at the 30\% level) because of the large
right-handed 2-3 mixing angle in the down quark Yukawa matrix.  This
tuning is so mild, we will not let it concern us further.  However, one 
should keep in mind that $\zeta$ may be
naturally close to one if the model is embedded into a larger
non-Abelian flavor group at a high scale.  Finally, we note that the
constraint on 2-3 mixing from $b\rightarrow s\gamma$ is very weak;
$(\delta_{23}^d)_{LL,RR}\sim {\cal O}(1)$ is
allowed~\cite{NPB477-321}.  We conclude that the non-Abelian model
presented in this section is consistent with the flavor changing
neutral current constraints.  Note that the model can be
extended trivially to the lepton sector by choosing the lepton 
transformation properties to be identical to those of the down quarks.  
Then the differences between the down quark and lepton masses can be
explained by fluctuations in the order one coefficients.

\begin{table}
\begin{center}
 \begin{tabular}{l|llllll}
 \hline\hline 
  {} &
  {$(\delta_{12}^d)_{LL}$} &
  {$(\delta_{13}^d)_{LL}$} &
  {$(\delta_{12}^u)_{LL}$} &
 \\ \hline 
  {Exp. upper bound} &
  {$4.0\times 10^{-2}$} &
  {$9.8\times 10^{-2}$} & 
  {$1.0\times 10^{-1}$} & 
 \\ \hline 
  {Prediction of the model} &
  {$\lambda^{5}$} & 
  {$\lambda^{3}$} & 
  {$\lambda^{6}$} & 
 \\
  {} &
  {$\sim 5.2\times 10^{-4}$} & 
  {$\sim 1.1\times 10^{-2}$} & 
  {$\sim 1.1\times 10^{-4}$} & 
 \\ \hline \hline
  {} &
  {$(\delta_{12}^d)_{RR}$} &
  {$(\delta_{13}^d)_{RR}$} &
  {$(\delta_{12}^u)_{RR}$} &
 \\ \hline 
  {Exp. upper bound} &
  {$4.0\times 10^{-2}$} &
  {$9.8\times 10^{-2}$} & 
  {$1.0\times 10^{-1}$} & 
 \\ \hline 
  {Prediction of the model} &
  {$(\zeta-1)\lambda^{2}$} & 
  {$(\zeta-1)\lambda^{3/2}$} & 
  {$\lambda^{6}$} & 
 \\
  {} &
  {$\sim 1.5\times 10^{-2}$} & 
  {$\sim 3.1\times 10^{-2}$} & 
  {$\sim 1.1\times 10^{-4}$} & 
 \\ \hline \hline
  {} &
  {$\overline{\delta}_{12}^d$} &
  {$\overline{\delta}_{13}^d$} &
  {$\overline{\delta}_{12}^u$} 
 \\ \hline 
  {Exp. upper bound} &
  {$2.8\times 10^{-3}$} & 
  {$1.8\times 10^{-2}$} & 
  {$1.7\times 10^{-2}$}
 \\ \hline 
  {Prediction of the model} &
  {$\sqrt{\zeta-1}\lambda^{7/2}$} & 
  {$\sqrt{\zeta-1}\lambda^{9/4}$} & 
  {$\lambda^{6}$}
 \\
  {} &
  {$\sim 2.7 \times 10^{-3}$} & 
  {$\sim 1.8 \times 10^{-2}$} & 
  {$\sim 1.1 \times 10^{-4}$}
 \\ \hline\hline
 \end{tabular} 
 \caption{Upper bounds on $(\delta_{ij}^q)_{LL,RR}$ and
$\overline{\delta}_{ij}^q$~\protect{\cite{NPB477-321}}.  Here, we take
all the squark and gluino masses to be 500 GeV. For comparison, we also
show the prediction of our model with $\lambda =0.22$ and 
$\zeta=c''_3/c''_0=1.3$.}
 \label{table:delta}
 \end{center}
\end{table}

\section{Abelian Model} \label{sec:abel} \setcounter{equation}{0}

We have seen that it is possible to construct models with 
non-Abelian flavor group factors in which flavor symmetries are 
broken via the dynamics of confinement.  Non-Abelian theories greatly 
alleviate the supersymmetric flavor-changing problem by imposing a 
natural degeneracy between the first two generation scalar masses in 
the flavor symmetric limit.  In this section, we show that models based 
on Abelian flavor symmetries can also incorporate our mechanism.  Such 
models solve the supersymmetric flavor problem by arranging an alignment 
of the quark and squark mass matrices, so that the squark masses are 
nearly diagonal in the quark mass eigenstate basis. The alignment is
strongest in the down quark sector, where the phenomenological constraints
are most powerful, and the Cabibbo angle originates in the up quark
sector. In this section, we will not present an exhaustive phenomenological 
analysis, but simply show that our symmetry-breaking mechanism can be 
combined with the prototypical alignment models of Nir and Seiberg \cite{NS}.

The important feature of the models of Ref.~\cite{NS}, as well
as similar models in Ref.~\cite{align}, is the presence of
two flavon fields, that transform under two independent U(1)
flavor symmetries: $S_1(-1,0)$ and $S_2(0,-1)$.  These
fields are assumed to acquire vevs
\begin{equation}
\langle S_1 \rangle = \epsilon_1 \sim \lambda^2
\,\,\,\,\,\,\,\, \mbox{ and } \,\,\,\,\,\,\,\,
\langle S_2 \rangle= \epsilon_2 \sim \lambda^3 \,\,\, .
\label{eq:svevs}
\end{equation}
Two models using these flavons are presented in Ref.~\cite{NS} (models A 
and B) which differ only in the flavor quantum number assignments of 
the matter fields.  We will explicitly consider model A below.   

The most elegant way of embedding this flavor sector into the
SU(3) theory described in the previous section is to choose
$S_1$ to be one of the meson fields, and $S_2$ to be one of the
baryons.  Since the baryon $S_2$ has one additional preon, its
symmetry breaking effect will be suppressed relative to the
meson $S_1$ by one factor of $\Lambda/M_*$.  If we again choose
this ratio to be the Cabibbo angle $\lambda$, we account
for Eq.~(\ref{eq:svevs}) in a natural way.  The two U(1) factors
can be taken such that
\begin{eqnarray}
Q_I = \sqrt{3} \,  Q_{\overline{p}} \nonumber \\
Q_{II} = - Q_B 
\end{eqnarray}
where the charges $Q_{\overline{p}}$ and $Q_B$ are defined as in the 
previous section. We can then make the identification
\begin{eqnarray}
S_1(-1,0) \equiv \sigma \nonumber \\
S_2(0,-1) \equiv B 
\end{eqnarray}
The remaining composites $S$, $A$, $\phi$, $\tilde{\phi}$
and $\bar{B}$, also have flavor quantum numbers, and may alter the Yukawa 
matrices slightly from the form presented in Ref.~\cite{NS}.   However, 
we will now show that the quark-squark alignment remains unaffected.  We 
will assume that the flavor SU(2) of the previous section is a good flavor 
symmetry (even though the matter fields are SU(2) singlets). Since 
the matter fields will have integral charges under the two U(1) factors, the 
lowest order combinations of the remaining composites that can contribute 
to the Yukawa textures are:
\begin{equation}
\bar{B}(0,+1) \sim \epsilon_2 \,\,\,\,\, \mbox{ and } \,\,\,\,\, 
A^2 (+1,0) \sim \epsilon_1^2
\end{equation}
Note that we have neglected terms involving $S$ which 
does not acquire a vev at lowest order.  The combination 
$(\phi \tilde{\phi})^2\sim (-1,0)$ couples in the same way as 
$S_1$, but is of higher order in $1/M_*$ and can also be neglected.
In model A, the matter fields are assigned charges
\begin{eqnarray}
Q_1(3,-1,+) & Q_2(1,0,-) & Q_3(0,0,+) \nonumber \\
U_1(-3,3,+) & U_2(-1,1,+) & U_3(0,0,+) \nonumber \\
D_1(-3,3,+) & D_2(1,0,-) & D_3(1,0,-) 
\end{eqnarray}
where the third entry is the charge under our anomalous $Z_2$
factor, defined in the previous section.  The original
textures for Model A in Ref.~\cite{NS}
\begin{equation}
h_u \sim \left(\begin{array}{ccc}
\epsilon_2^2 & \epsilon_1^2 & 0 \\
0 & \epsilon_2 & \epsilon_1 \\
0 & 0 & 1 \end{array}\right)      \,\,\,\,\,\,\,\,\,\,\,
h_d \sim \left(\begin{array}{ccc}
\epsilon_2^2 & 0 & 0 \\
0 & \epsilon_1^2 & \epsilon_1^2 \\
0 & \epsilon_1 & \epsilon_1 \end{array}\right)
\end{equation}
become
\begin{equation}
h_u \sim \left(\begin{array}{ccc}
\epsilon_2^2 & \epsilon_1^2 & \epsilon_1^6 \epsilon_2 \\
\epsilon_1^4\epsilon_2^3 & \epsilon_2 & \epsilon_1 \\
\epsilon_1^9 \epsilon_2^3 & \epsilon_1^5\epsilon_2 & 1 
\end{array}\right)      \,\,\,\,\,\,\,\,\,\,\,
h_d \sim \left(\begin{array}{ccc}
\epsilon_2^2 & \epsilon_1^7\epsilon_2 & \epsilon_1^7\epsilon_2 \\
\epsilon_1^4\epsilon_2^3 & \epsilon_1^2 & \epsilon_1^2 \\
\epsilon_1^9 \epsilon_2^3  & \epsilon_1 & \epsilon_1 \end{array}\right)  
\,\,\, .
\label{eq:newent}
\end{equation}
The scalar mass matrices are not holomorphic functions of the flavon
fields, so their textures remain unchanged:
\[
\frac{(\tilde{M^2_q})^0_{LL}}{\tilde{m}^2} \sim \left(\begin{array}{ccc}
1 & \epsilon_1^2 \epsilon_2 & \epsilon_1^3 \epsilon_2 \\
\epsilon_1^2 \epsilon_2  & 1 & \epsilon_1 \\
\epsilon_1^3 \epsilon_2& \epsilon_1 & 1 \end{array}\right)
\,\,\,\,\,\,\,\,\,\,\,
\frac{(\tilde{M^2_u})^0_{RR}}{\tilde{m}^2}  \sim \left(\begin{array}{ccc}
1 & \epsilon_1^2\epsilon_2^2 & \epsilon_1^3\epsilon_2^3 \\
\epsilon_1^2\epsilon_2^2 & 1 & \epsilon_1 \epsilon_2 \\
\epsilon_1^3\epsilon_2^3 & \epsilon_1\epsilon_2 & 1 \end{array}\right)  
\]\begin{equation}
\frac{(\tilde{M^2_d})^0_{RR}}{\tilde{m}^2}  \sim \left(\begin{array}{ccc}
1 & \epsilon_1^4\epsilon_2^3 & \epsilon_1^4\epsilon_2^3 \\
\epsilon_1^4\epsilon_2^3 & 1 & 1 \\
\epsilon_1^4\epsilon_2^3 & 1 & 1 \end{array}\right) \,\,\, . 
\end{equation}
If we now go to the quark mass eigenstate basis, all the rotations on 
the left-handed quark fields that are induced by the additional entries 
in Eq.~(\ref{eq:newent}) do not alter the order of magnitude of any 
off-diagonal squark mass matrix elements.  Only the 1-2 rotation in the 
right-handed down sector is large enough to change the (1,2) entry of
$(\tilde{M^2_d})_{RR}$ from $\epsilon_1^4\epsilon_2^3$ to 
$\epsilon_1^2 \epsilon_2^3 \sim 10^{-9}$.  The bound on this entry 
from flavor changing neutral currents is of order $10^{-2}$, and is still 
easily satisfied in the modified model.

Thus, the presence of additional flavons implied by our symmetry-breaking
mechanism does not disturb the quark-squark alignment.  One can easily 
verify that the same is true for Model B of Ref.~\cite{NS} as well.

\section{Conclusions}

Supersymmetric theories have two sets of small dimensionless flavor
parameters: one describes the quark and lepton mass ratios and mixing
angles, while the other describes squark and slepton non-degeneracies
and mixings, which are constrained from flavor-changing processes.  We 
have described a general framework of theories with a flavor symmetry, and 
given two explicit realistic models, where
\begin{itemize}
\item Flavor symmetry breaking is forced by strong supersymmetric gauge
interactions.
\item All non-zero vevs have a magnitude of order the $\Lambda$ parameter of
the new strong gauge force. All flavor symmetry breaking occurs at a single
scale, and there is a single small dimensionless parameter, $\Lambda/M_*$,
where $M_*$ is the cutoff for the theory.
\item The flavor symmetry allows certain higher dimension F and D operators
coupling quarks ($q$), Higgs ($H$) and preons ($p$),
\[
[\bar{q} q H (p/M_*)^n]_F \,\,\,\,\, \mbox{ and } 
\,\,\,\,\,
[q^\dagger q  (p/M_*)^m]_D \,\,\, , 
\]
generating small
entries in the quark and squark mass matrices of order $(\Lambda / M_*)^n$ and 
$(\Lambda / M_*)^m$ respectively.
\item While flavor symmetry breaking is forced, there is a large vacuum
degeneracy --- the quark and squark mass matrices are functions on moduli
space. This degeneracy can be lifted in a favorable direction by the combined
use of the $X$ fields and soft, positive supersymmetry breaking squared
masses.
\end{itemize}

\begin{center}               
{\bf Acknowledgments} 
\end{center}
We would like to thank N. Arkani-Hamed, H. Murayama and J. Terning for 
useful discussions. This work was supported in part by the Director, Office 
of Energy Research, Office of High Energy and Nuclear Physics, Division of 
High Energy Physics of the U.S. Department of Energy under Contract
DE-AC03-76SF00098.  LJH was also supported in part by the National
Science Foundation under grant PHY-95-14797.

%\appendix
%\section{Other Models}
%\setcounter{equation}{0}


\begin{thebibliography}{99} 
\frenchspacing

\bibitem{Georgi}
See, for example, H. Georgi, {\em Weak Interactions and Modern Particle
Theory}, Benjamin/Cummings, Menlo Park, (1984), and references therein.
\bibitem{Higgs}
P.W. Higgs, {\sl Phys. Lett.} \/ {\bf 12}, 132 (1964); {\sl Phys. Rev. Lett.} 
\/ {\bf 13}, 508 (1964); {\sl Phys. Rev.} \/ {\bf 145}, 1156 (1966);
F. Englert and R. Brout, {\sl Phys. Rev. Lett.} \/ {\bf 13}, 321 (1964);
T.W.B. Kibble, {\sl Phys. Rev.} \/ {\bf 155}, 1554 (1967).

\bibitem{RadiativeBreaking}
L. Ibanez, G.G. Ross, {\sl Phys .Lett.} \/ {\bf B110}, 215 (1982);
K. Inoue, A. Kakuto, H. Komatsu, and S. Takeshita,
{\sl Prog. Theor. Phys.} \/ {\bf 68} 927 (1982).

\bibitem{seiberg}
N.~Seiberg, {\sl Phys. Rev.} \/ {\bf D49}, 6857 (1994).

\bibitem{cheng}
H.-C. Cheng, FERMILAB-PUB-97-019-T, Jan 1997, 
hep-ph/9702214.

\bibitem{ads}
I.~Affleck, M.~Dine, and N.~ Seiberg, {\sl Nucl. Phys.} \/  
{\bf B256}, 557 (1985).

\bibitem{peskin}
O.~Aharony, J.~Sonnenschein, M.E.~Peskin, S.~Yankielowicz,
{\sl Phys. Rev.} \/ {\bf D52}, 6157 (1995).

\bibitem{dlk}
M. Dine, R. Leigh, and A. Kagan, 
{\sl Phys .Rev.} \/ {\bf D48}, 4269 (1993).

\bibitem{lattice}
R.~Gupta and T.~Bhattacharya, {\sl Phys. Rev.} \/ {\bf D55}, 7203 (1997).

\bibitem{NPB477-321} 
See, for example, F.~Gabbiani, E.~Gabrielli, A.~Masiero, and
L.~Silvestrini, {\sl Nucl. Phys.} \/ {\bf B477}, 321 (1996).

\bibitem{NS}
Y.~Nir and N.~Seiberg, {\sl Phys. Lett.} \/ {\bf B309},
337 (1993)  

\bibitem{align}
M.~Leurer, Y.~Nir, and N.~Seiberg, 
{\sl Nucl. Phys.} \/ {\bf B398}, 319 (1993); 
M.~Leurer and N.~Seiberg, {\sl Nucl. Phys.} \/ {\bf B420}, 
468 (1994).

\end{thebibliography}
\end{document}